\documentclass[a4paper, twocolumn, twocolappendix]{aastex631}
\usepackage{mathrsfs}
\usepackage{amsmath}
\numberwithin{equation}{section}
\usepackage{graphicx}
\graphicspath{{images/}}
\usepackage{times}
\usepackage[varg]{txfonts}
\usepackage{hyperref}

\begin{document}
\title{Could the high-mass black holes from gravitational-wave observations be explained by lensing?}

\author[0000-0002-9115-0287]{Ritesh Harshe}
\affiliation{International Centre for Theoretical Sciences, Tata Institute of Fundamental Research, Bangalore 560089, India}

\author[0000-0002-6602-3913]{R. Prasad}
\affiliation{International Centre for Theoretical Sciences, Tata Institute of Fundamental Research, Bangalore 560089, India}

\author[0000-0001-7519-2439]{Parameswaran Ajith}
\affiliation{International Centre for Theoretical Sciences, Tata Institute of Fundamental Research, Bangalore 560089, India}

\begin{abstract}
The high-mass ($M \gtrsim 30 M_\odot$) black holes (BHs) from the gravitational-wave (GW) observations of LIGO and Virgo came as a surprise to many astronomers. While the collapse of metal-poor massive stars could produce such BHs, gravitational lensing has been invoked to explain their high masses. Broadhurst, Diego, and Smoot (henceforth BDS) argued that the mass distribution of BHs in coalescing binaries is very similar to that of the galactic BHs, and the inferred high masses are the result of neglecting the lensing magnification. They also proposed a redshift distribution of binary BH (BBH) mergers to explain the observed LIGO-Virgo mass distribution. We ask whether such a model is consistent with different aspects of the GW observations: 1) the observed number of BBH mergers, 2) the distribution of their redshifted total mass and apparent luminosity distance, 3) the non-detection of strongly lensed events, and 4) the non-observation of the stochastic GW background. By simulating lensed BBH mergers with the BDS model and comparing them with observations, we conclude that no choice of BDS model parameters is consistent with all aspects of the observations. Lensing magnification is not a viable explanation for the high-mass BHs discovered by LIGO and Virgo.
\end{abstract}

%% https://astrothesaurus.org
\keywords{Black holes (162) --- Gravitational lensing (670) --- Gravitational waves (678)}

%%natbib \citep \citet
\section{Introduction} \label{sec:intro}

Gravitational-wave (GW) observations by LIGO~\citep{LIGOScientific:2014pky}, Virgo~\citep{VIRGO:2014yos}, and KAGRA~\citep{KAGRA:2020tym} (LVK) have uncovered a new population of extragalactic high-mass ($M\gtrsim30 M_\odot$) black holes (BHs)~\citep{LIGOScientific:2018mvr, LIGOScientific:2020ibl, KAGRA:2021vkt, LIGOScientific:2025slb}. In contrast, most of the BHs from galactic X-ray observations are significantly less massive ($M\lesssim10M_\odot$; ~\cite{Corral-Santana:2015fud}). The conventional explanation is that these massive BHs are remnants of high-mass stars in low-metallicity environments~\citep{Belczynski:2009xy, Fryer:2011cx}. Another possibility is that they are the much-speculated primordial BHs~\citep{Carr:1974nx}.  

\cite{Broadhurst:2018saj} and \cite{Diego:2021fyd} (henceforth, BDS) argued that gravitational lensing of GWs could be the possible reason for the apparent abundance of these high-mass BHs. It is well known that lensing magnification can bias the distance and mass estimation of GW sources~\citep{Dai:2016igl}. BDS argue that the true mass distribution is consistent with that of the galactic BHs, and the observed high-mass population is an artifact caused by a large fraction of lensed GW signals in the LVK data. They assume that the BH mass distribution across the universe is the same as the galactic distribution, and propose a merger rate model that favors a larger number of mergers occurring at higher redshifts as compared to conventional rates. They argue that such a model will yield the observed mass distribution found in the LVK data.

While an intriguing possibility, most of the conventional astrophysical models predict only a much smaller fraction of the binary BH (BBH) events to be lensed~\citep{Ng:2017yiu, Li:2018prc, Oguri:2018muv, Xu:2021bfn, Wierda:2021upe, Mukherjee:2021qam, Barsode:2024zwv}. Further, \cite{Farah:2025ews} have pointed out the pitfalls of identifying the high-mass events as lensed ones. Most studies on GW lensing acknowledge that, although this explanation for the massive BHs is unlikely to hold, such a possibility cannot be ruled out~\citep{Smith:2017mqu, LIGOScientific:2021izm, LIGOScientific:2023bwz}. In this \emph{Letter}, we examine whether such a model is consistent with all the observed features of data, namely, the observed number of BBH mergers, the observed distribution of redshifted total mass and apparent luminosity distance, 
 the non-detection of strongly lensed events, 
and the non-observation of the stochastic GW background. By performing astrophysical simulations of lensed BBH sources using the BDS model, we conclude that no parameter choice in the model is consistent with all observed features, and hence the model is untenable. Lensing cannot explain the high-mass BHs from LVK observations.

\section{Methods} \label{sec: methods}

\subsection{Lensing Magnification}\label{subsec: grav_lensing}
In gravitational lensing by galaxies and clusters, wavelengths of the GWs are much smaller than the lens size (geometric optics approximation), which are, in turn, much smaller than the cosmological distances (thin lens approximation)~\citep{schneiderbook}. In the geometric optics regime, lensing is achromatic. Lensing endows a GW signal $h(f)$, written in the frequency domain, with a magnification $\mu$, time delay $\delta t$, and a constant phase shift $\delta \phi_0$~\footnote{The time delay (as compared to the unlensed signal) is caused by the geometric path differences and the gravitational potential of the lens. The phase shift ($0, \pi/2$ or $\pi$) depends on whether the image is formed at the minimum, saddle or maximum of the time delay surface.}.
\begin{equation}
    h^\ell(f) = \mu^{1/2} \, h(f) \, e^{i [2 \pi f \delta t + \delta \phi_0]}. 
\end{equation}
The lensing magnification is fully degenerate with the luminosity distance $d_L$ to the source, which will bias our inference of $d_L$. Also, limiting to the dominant (quadrupole) mode GW signals, $\delta t$ and $\delta \phi_0$ are degenerate with the coalescence time and coalescence phase of the GW signal, and are not directly measurable. 

If the GW is sufficiently away from the lens, only a single magnified image is formed (weak lensing). However, when a GW passes sufficiently close to a lens, it can form multiple images (strong lensing). These multiple images, indexed by~$j$, are endowed with different values of $\mu, \delta t$ and $\delta \phi_0$ 
\begin{equation}
    h^\ell_j(f) = \mu^{1/2}_j \, h(f) \, e^{i [2 \pi f \delta t_j + \delta \phi_{0j}]}. 
\end{equation}
If multiple images are detected, this would allow us to measure the magnification ratio $\mu_r$ as well as the relative time delay $\Delta t$ and phase shift $\Delta \phi$ between the lensed images. 

GWs are redshifted due to cosmological expansion. Since GW signals from compact binaries are fully scalable with the total mass $M$ of the system, the cosmological redshift $z$ is degenerate with $M$ --- what we can infer is the ``redshifted'' mass $M^z \equiv M (1 + z)$. Since compact binaries are standard sirens, we can measure the luminosity distance $d_L$ \citep{Schutz:1986gp, Holz:2005df}. If we assume a cosmological model, we can infer the redshift $z$ corresponding to the luminosity distance $d_L$, that is, $z(d_L)$, which allows us to infer the true (source frame) mass
\begin{equation}
    M = M^z~\left[1 + z\left(d_L \right)\right]^{-1}. 
\end{equation}
When a GW signal is lensed with an unknown magnification $\mu$, instead of the true luminosity distance $d_L$, we will infer only the apparent luminosity distance $\tilde{d_L} \equiv d_L/\mu^{1/2}$.  As a result, the estimated value of source frame mass will also be biased 
\begin{equation}
    M^\mathrm{b} = M^z~\left[1 + z \left(\tilde{d_L} \right)\right]^{-1}. 
\end{equation}
Redshifted mass is unaffected by lensing, and the mass ratio $q \equiv m_1/m_2$ of the binary is affected neither by the cosmological expansion nor by lensing. This means that the inferred values of the component masses will also be biased due to lensing,
\begin{equation}
    m^\mathrm{b}_{1,2} = m^z_{1,2}~\left[1 + z \left(\tilde{d_L} \right)\right]^{-1}. 
\end{equation}

A magnified ($\mu > 1$) GW signal will cause the inferred distances to bias towards smaller values, and hence the inferred masses to bias towards larger values. BDS argued that the apparent high masses of the BHs inferred from LVK observations are an artifact of this lensing magnification. 

\subsection{Simulating BBHs Using the BDS model}\label{subsec: inj_gen}
The BDS model assumes that the mass distribution of BHs in coalescing binaries is the same as the galactic BH mass distribution, inferred from X-ray binaries in the~\cite{Corral-Santana:2015fud} catalog: 
\begin{equation}
    \label{eq:log-normal-mass-distr}
    \frac{dN}{dm} = \frac{1}{m~\sigma_m}~\exp\left({\frac{-(\log_{10} m - \mu_m)^2}{2\sigma_m^2}}\right),
\end{equation}
where $\mu_m=\log_{10}(8)$ and $\sigma_m = 0.27 \, M_\odot$. \cite{Diego:2021fyd}) also proposed the following BBH merger rate (BBH mergers per unit time per unit comoving volume in the source frame):
\begin{equation}
    \label{eq:rate-of-mergers}
    R(z) = 
    \begin{cases}
        A_1\exp\left({\dfrac{t(z)-t(z_\mathrm{peak})}{t_h}}\right) & z<z_\mathrm{peak},\\
        A_2\dfrac{(1+z)^{2.7}}{1+ \left(\dfrac{1+z}{2.9} \right)^{5.6}} & \mathrm{otherwise},
    \end{cases}
\end{equation}
where $t(z)$ is the lookback time in Gyr at redshift $z$, the value of $z_\mathrm{peak} = 1.8$ and $A_2 = A_1 [1+((1+z_\mathrm{peak})/2.9)^{5.6}]/(1+z_\mathrm{peak})^{2.7}$. The parameter $A_1$ represents the peak merger rate and corresponds to the value at $z_\mathrm{peak}$, while $t_h$ characterizes the fall of the merger rate from its maximum value to the present epoch. BDS argued that $A_1 \approx 30000 \, \mathrm{yr}^{-1} \, \mathrm{Gpc}^{-3}$ and $t_h = 1.25 \, \mathrm{Gyr}$ reproduce the observed distribution of BH masses in LVK data.

\begin{figure}
    \centering
    \includegraphics[width=\linewidth]{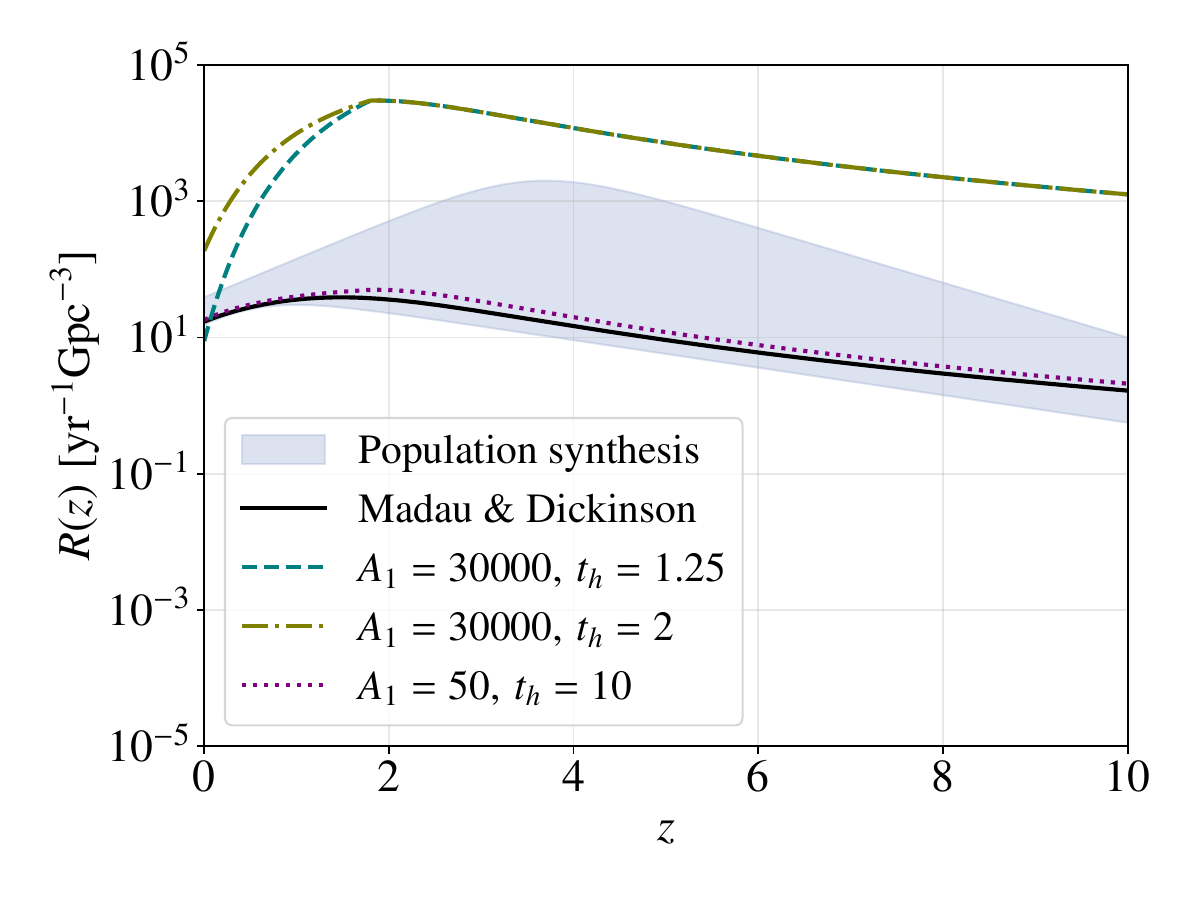}
    \caption{Dotted-dashed lines show the BDS merger rate $R(z)$ for different values of the model parameters $A_1$ in units of $\mathrm{yr}^{-1}\, \mathrm{Gpc}^{-3}$ and $t_h$ in units of $\mathrm{Gyr}$ (shown in legends). In contrast, the shaded band shows the merger rates predicted by various population synthesis models in the literature. The black curve corresponds to the same obtained from the star formation model of~\cite{Madau:2014bja} that is scaled with the local merger rate inferred from LVK observations. 
    }
    \label{fig:bds-merger-rate}
\end{figure}

Figure~\ref{fig:bds-merger-rate} shows the merger rate $R(z)$ for a few choices of $A_1$ and $t_h$. In comparison, the merger rate model inspired by star formation history~\citep{Madau:2014bja} matched to low-redshift GW observations from the first three observing runs of LVK is also plotted (black solid curve). The shaded band covers the merger rates predicted by various population synthesis models~\citep{Belczynski:2005mr, Belczynski:2009xy, Dominik:2013tma, Marchant:2018kun, Boco:2019teq, Eldridge:2018nop, Neijssel:2019irh, Santoliquido:2020axb}. This is made using the fits given by Eqs. A1 and A2 in \cite{LIGOScientific:2021izm}.

We simulate a population of BBHs following the BDS model. For each choice of the model parameters $A_1$ and $t_h$, the component masses are sampled from the log-normal distribution, Eq.~\eqref{eq:log-normal-mass-distr}, within the range $m_1, \, m_2 \in [5, 100]\, M_\odot$ while the source redshifts are sampled from redshift distributions derived from merger rates, Eq.~\eqref{eq:rate-of-mergers}, within the range $ z_{s} \in [0, 10]$. We assume that the spins are aligned to the orbital angular momentum, with magnitudes uniformly distributed within the range ${a_1, \, a_2 \in [0, \, 0.99]}$. Binaries are distributed uniformly in the sky with isotropic orientations. GW signals are modeled using the \textsc{IMRPhenomXPHM} waveform approximant~\citep{Pratten:2020ceb}~\footnote{We have also compared these SNRs with those computed using the \textsc{NRSur7dq4} approximant~\citep{Varma:2019csw}. For the vast majority of the parameter space, the fractional difference in the SNRs is within a few percent. Even for the worst case scenario (binaries with high masses, high mass ratios and high spins), the difference is within $\sim 15\%$. Note that we neglected the effect of spin precession, as the dominant spin effects in the waveform are determined by the effective spin parameter (see, e.g.,~\cite{Ajith:2009bn}).}.

We consider multiple observing runs of LVK (O1, O2, O3a, and O3b). The arrival times for events in the LVK observing runs are uniformly sampled with the corresponding start and end times of each run. To compute the {expected} number of detectable signals as well as the signal-to-noise ratio (SNR) of the stochastic GW background (SGWB), we multiply the observation time with the following duty cycle of the LIGO-Virgo detectors: 0.48 for O1, 0.46 for O2, 0.82 for O3a, and 0.85 for O3b~\citep{ligo2016o1, LIGOScientific:2020ibl, KAGRA:2023pio}. We use representative noise power spectral densities from each observing run~\citep{LIGO-G1600150, LIGO-G1600151, LIGO-G1801950, LIGO-G1801952, LIGO-T2000012}. Throughout this work, we assume a flat $\Lambda$CDM cosmology with parameters from Planck 2018 results~\citep{Planck:2018vyg}.

\subsection{Lensing Scheme}\label{subsec: lensing_scheme}

Each lens is approximated as a singular isothermal sphere (SIS), which provides a reasonable model for galaxy-scale lenses. A source is classified as strongly lensed if it lies within the Einstein radius $r_E$ of the SIS lens. It is convenient to define a dimensionless source position $y$ on the lens plane (in units of $r_E$). When $y \leq 1$ (strong lensing regime), two images are produced with magnifications $\mu_{\pm} = 1 \pm 1/y$. For sources with $y > 1$ (weak lensing regime), only a single image is formed, with magnification $\mu_+$. We sample $y$ from the probability density given by~\cite{Barsode:2024wda},
\begin{equation}
\frac{dp}{dy}(z_s) = 2\tau(z_s) \, y \, \exp[-y^2\tau(z_s)],
\label{eq:dPdy}
\end{equation}
where $\tau(z_s)$ denotes the lensing optical depth at the source redshift $z_s$, computed following the prescription in~\cite{Jana:2022shb} with \cite{Behroozi:2012iw} halo mass function model; more details are available in Appendix~\ref{appendix: choice_of_the_lens_model}. Note that every GW signal is lensed; however, most sources are located far away from the lenses ($y \gg 1$), and therefore, their lensing magnification is close to unity ($\mu \simeq 1$). 

A weakly lensed event ($y > 1$) produces a single image with magnification $\mu$ and arrival time $t$. This image is deemed detected when its optimal SNR $\rho^\ell = \mu^{1/2}\rho(t) \geq 8$, where $\rho(t)$ is the optimal SNR of the corresponding unlensed signal. A strongly lensed event ($y \leq 1$) produces multiple images with magnifications $\mu_i$ arriving at times $t_i$. The SNR of the $i$th image is given by $\rho^\ell_i = |\mu_i|^{1/2} \, \rho(t_i)$. Since the two images arrive at different times, note that even their unlensed SNRs will be, in general, different. If only one of the $\rho^\ell_i \geq 8$, then the event is considered a detectable weakly lensed event. If both the $\rho^\ell_i \geq 8$, then it is considered a detectable strongly lensed one. 

\subsection{Observables}

The following are the observables used to constrain the BDS model parameters $A_1$ and $t_h$.

\subsubsection{Number of Detectable BBH mergers}\label{subsec: detectable_event_methods}
The number of detectable BBH mergers is sensitive to the distribution of source mass and the merger rate, making it a useful probe for constraining the parameters $A_1$ and $t_h$. We simulate the number of BBH mergers in each observing scenario, 
\begin{equation}
\Lambda_\mathrm{sim}(A_1, t_h)=T_{\mathrm{obs}} \int_0^{10} \frac{dz}{1+z} R(z; A_1, t_h) \frac{dV_c}{dz},
\label{eq:num_bbh_mergers}
\end{equation}
where $T_{\mathrm{obs}}$ is the observation time, $R(z; A_1, t_h)$ is given by Eq.~\eqref{eq:rate-of-mergers}, $dV_c/dz$ is the differential comoving volume, while the factor $1/(1+z)$ is due to the cosmological time dilation. The redshift distribution of the mergers is given by the integrand of Eq.~\eqref{eq:num_bbh_mergers}. Then, using the lensing scheme in Sec.~\ref{subsec: lensing_scheme}, the number of detectable events is computed (events crossing an optimal SNR threshold of 8 and which arrive within O1, O2, or O3 runs\footnote{Note that the detectability condition used in actual searches is much more complex, due to the non-Gaussian tails in the data. However, the Poisson uncertainties on the number of detectable events that we consider takes care of this difference.}). Detectable events from different observing runs are summed to obtain the total expected number of detectable events, $\Lambda(A_1, t_h)$.  We assume the detection of GW events as a Poisson process with mean $\Lambda$. We compare the predicted value of $\Lambda(A_1, t_h)$ with $3\sigma$ lower/upper bounds on $\Lambda$ estimated from the actual number of observed events ($N_\mathrm{det} = 83$) in the LVK runs~(see Appendix~\ref{appendix:detectable_events_and _strongly_lensed_pairs_thresholds}). This allows us to constrain the values of $A_1$ and $t_h$ so that the number of detectable BBH mergers predicted by the model is consistent with the observations.

\subsubsection{Detectable Fraction of Strongly Lensed Pairs}\label{subsec: strongly_lensed_pairs}

The fraction of strongly lensed events depends on the merger rate distribution, since the strong lensing optical depth $\tau(z_s)$ depends on the source redshift $z_s$. Therefore, the observed fraction $u$ of strongly lensed events helps us constrain  $t_h$ (the lensing \emph{fraction} is independent of $A_1$). Note that $u$ includes only events in which both images exceed the detection SNR threshold and arrive within O1, O2, or O3; if only one image is detectable, it is excluded from the strongly lensed pair count. We compute the expected value of $u$ as a function of $t_h$, and compare it with the posterior of $u$ obtained from O1-O2-O3 data to constrain $t_h$. This is done using a Bayesian likelihood analysis described below. 

Multiple copies of GW signals produced by strong lensing will have the same shape, except for an overall magnification and time/phase delay (see Sec.~\ref{subsec: grav_lensing}). This means that the posterior distributions of the parameters $m_1^z, \, m_2^z, \, a_1, \, a_2,$ sky location, orientation, etc (but not $d_L$), will be consistent between the two events. \cite{Haris:2018vmn} developed a \emph{Posterior Overlap} statistic $B^\mathrm{ovlp}$ that tests the consistency of the two posteriors, which has been used to search for strongly lensed events in O1-O2-O3 data~\citep{Hannuksela:2019kle,LIGOScientific:2021izm, LIGOScientific:2023bwz}. We take the list of $B^\mathrm{ovlp}$ values obtained from pairs of events from O1-O2-O3 data and compute its likelihood to follow a distribution that is a mixture of $B^\mathrm{ovlp}$'s from lensed and unlensed pairs. 
\begin{equation}
    \label{eq:blu_likelihood_u}
    P(B^\mathrm{ovlp}|u) = \frac{N_p^\ell \, P(B^\mathrm{ovlp}|H_L) \, +  (N_p - N_p^\ell) \, P(B^\mathrm{ovlp}|H_U)}{N_p}.
\end{equation}
Above, $P(B^\mathrm{ovlp} | H_L)$ and $P(B^\mathrm{ovlp} | H_U)$ are the expected distributions of $B^\mathrm{ovlp}$ from lensed and unlensed pairs, respectively. Also, $N_p$ is the total number of pairs that we consider~\footnote{If we consider all possible pairs, there will be $N_\mathrm{det}(N_\mathrm{det}-1)/2$ of them. However, we consider only event pairs from the same observing runs (i.e., O1-O1, O2-O2, O3a-O3a, and O3b-O3b), as presently searches are done only for these pairs. Hence, $N_p = 1521$ in our calculations.}, and $N_p^\ell = u \, N_\mathrm{det} $ are the lensed pairs among them. Naturally, the number of unlensed pairs is $N_p - N_p^\ell$.

We can evaluate the likelihood function Eq.~\eqref{eq:blu_likelihood_u} on the list of posterior overlap values $\{B^\mathrm{ovlp}_i\}$ that we obtain from the O1-O2-O3 data. Assuming a flat prior on $u$, the posterior on $u$ from the full pair of events is.
\begin{equation}
    \label{eq:observed_lensing_fraction_posterior}
    P(u|\{B^\mathrm{ovlp}_i\}) \propto \prod_{i=1}^{N_p} P(B^\mathrm{ovlp}_i|u).
\end{equation}
We compute the $3\sigma$ upper limit on $u$ (see Appendix \ref{appendix:on_the_non_detection_of_strongly_lensed_pairs_until_o3b}). Any value of $t_h$ that predicts a lensing fraction larger than this upper limit can be ruled out~{\footnote{These constraints can be further tightened making use of the non-observation of sub-threshold lensed signals; see, e.g.~\cite{LIGOScientific:2023bwz}.}. 

Note that our analysis is independent of the time delay distribution. BDS criticized the LVK analysis for penalizing event pairs with large time delays, since it used a galaxy-lens-derived time delay distribution to identify lensed events. Our analysis avoids this by relying solely on source-parameter consistency, making it independent of lens population assumptions. 

\subsubsection{Distribution of Total Redshifted Mass and Apparent Luminosity Distance}~\label{subsec: mass_dL_distr}
The redshifted total mass $M^z \equiv M (1+z)$ and the apparent luminosity distance $\tilde{d_L} \equiv d_L/\mu^{1/2}$ are estimated without any bias from lensed GW signals. We compute the joint distribution of $\theta \equiv \{M^z, \tilde{d_L}\}$ predicted by the BDS model for different values of $t_h$ and compute its consistency with the posteriors of $\theta$ estimated from O1-O2-O3 events \citep{ligo_virgo_kagra_2021_gwtc21_pe, ligo_virgo_kagra_2021_gwtc3_pe, LIGOScientific:2021usb, KAGRA:2021vkt}. This is done using a hierarchical population inference~\citep{Thrane:2018qnx, Mandel:2018mve}.   

Note that the normalized distribution of $\theta$ is independent of $A_1$. The marginal likelihood of a set of events $\{d\}$, given $t_h$ is
\begin{equation}
    P(\{d\} \, |\, t_h) \propto \frac{\Pi_{i=1}^{N_\mathrm{det}} \int d\theta ~ P(d_i|\theta) ~ P_\mathrm{pop}(\theta|t_h)}{ \left[ \int d\theta ~ P_\mathrm{det}(\theta) ~ P_\mathrm{pop}(\theta|t_h) \right]^{N_\mathrm{det}} }.
    \label{eq:th_likehood_from_mass_dist}
\end{equation}
Above, $P(d_i|\theta)$ is the likelihood of observing one event $d_i$ given the set of source parameters $\theta$,  $P_\mathrm{pop}(\theta|t_h)$ is the expected (prior) distribution of $\theta$ predicted by the BDS model, given a value of $t_h$, while $P_\mathrm{det}(\theta)$ is the detectability (fraction of detectable events) for a set of source parameters $\theta$. The likelihood $P(d_i|\theta)$ of individual events is obtained by dividing the corresponding posteriors $P(\theta|d_i)$ by the priors $P_\mathrm{PE}(\theta)$ used in the parameter estimation of individual events. Note that all the distributions are marginalized over all the other source parameters. We then obtain the posterior of $t_h$ using the Bayes theorem, assuming a flat prior on $t_h$. That is, 
\begin{equation}
    P(t_h  \, |\, \{d\} ) \propto P(\{d\} \, |\, t_h).
     \label{eq:th_posteriors_bayes_theorem}
\end{equation}

\subsubsection{Stochastic GW Background}\label{subsec: sgwb_methods}
The unresolved incoherent superposition of GWs emitted by multiple sources produces the SGWB. If the high-redshift merger rate is very high, the SGWB signal should already be detectable. Thus, the non-detection of the SGWB signal can be used to constrain the parameter space of the BDS model. 

The SGWB amplitude is given by~\citep{Phinney:2001di}
\begin{equation}
    \label{eq:sgwb-amplitude}
    \begin{split}
        \Omega_\mathrm{GW}(f) = \frac{f}{\rho_c c^2}\int_0^\infty dz &\int d\theta \frac{dV_c}{dz} p(\theta)\frac{R(z, A_1, t_h)}{1+z}\\
        &
        \left(\frac{1+z}{4\pi c d_L(z)^2}\frac{dE_\mathrm{GW}(\theta)}{df_r}\right)\Bigg |_{f_r=(1+z)f},
    \end{split}
\end{equation}
where $f$ is the observed frequency, $\rho_c$ is the critical density for a flat universe, $d_L$ is the luminosity distance to the sources, $dV_c/dz$ is the differential comoving volume, $p(\theta)$ is the probability distribution of source parameters, $R(z; A_1, t_h)$ is the BBH merger rate, $dE_\mathrm{GW}/df_r$ is the GW energy flux in the source frame per frequency bin, and $f_r = (1+z)f$ is the source frame frequency of GW emission. We compute $dE_\mathrm{GW}/df_r$ using the \textsc{IMRPhenomXP} waveform model~\citep{Pratten:2020fqn}.

The SNR of SGWB from a network of two detectors is given by~\citep{Allen:1997ad}
\begin{equation}
    \label{eq:sgwb-snr}
    \mathrm{SNR} = \frac{3H_0^2}{10\pi^2}\sqrt{2T_\mathrm{obs}}\left(\int_0^\infty df \sum_{i=1}^n \sum_{j>i}\frac{\gamma_{ij}^2(f)\Omega_\mathrm{GW}^2(f)}{f^6 P_i(f)P_j(f)}\right)^{1/2},
\end{equation}
where $H_0$ is the Hubble constant, $T_\mathrm{obs}$ is the effective observation time (after considering the duty cycle), $i, j$ runs over detector pairs, $P_i(f)$ and $P_j(f)$ are the one-sided noise power spectral densities of the two detectors, and $\gamma_{ij}(f)$ is the normalized isotropic overlap reduction function between the detector pair. Since the contribution of the H1-L1 pair to the SNR is significant due to their larger overlap function at lower frequencies compared to other pairs, we find SNR with the H1-L1 pair only~\citep{LIGOScientific:2017zlf}. We calculate the SNR for various $A_1, \, t_h$ values of the BDS model. The fact that we have not seen the SGWB enables us to rule out $A_1, \, t_h$ values that produce SNR exceeding $1$, $2$ or $3$, depending on the adopted confidence level~\citep{LIGOScientific:2017zlf, KAGRA:2021kbb}.

\section{Constraints on the BDS model} \label{sec: results_and_discussion}
\paragraph{Based on the Number of Detectable BBH mergers}~ 
Figure~\ref{fig:detectable_events} shows the total number of detectable events as a function of $A_1$ and $t_h$, following the approach described in Sec.~\ref{subsec: detectable_event_methods}. The number of detectable events increases with $A_1$, {due to an overall rise in the number of mergers}; the number of detections also increases with $t_h$, {owing to the larger contribution from lower-redshift mergers}. While the predicted numbers range from $\sim 1 - 10^4$, the $3\sigma$ credible range for the Poisson mean obtained from the data is $60 - 115$. This is obtained using the fact that LVK has observed $83$ BBH merger events till O3b~(\cite{KAGRA:2021vkt}; see Appendix~\ref{appendix:detectable_events_and _strongly_lensed_pairs_thresholds} for details). 
A large portion of the $A_1-t_h$ parameter space appears inconsistent with the observed number of detectable events, leaving only a small region bounded by the $\Lambda = 60$ and $115$ contours. The regions ruled out correspond to $A_1-t_h$ that predict either too many or too few detectable events to be consistent with the observed count.

\begin{figure}[t]
    \centering
    \includegraphics[width=\linewidth]{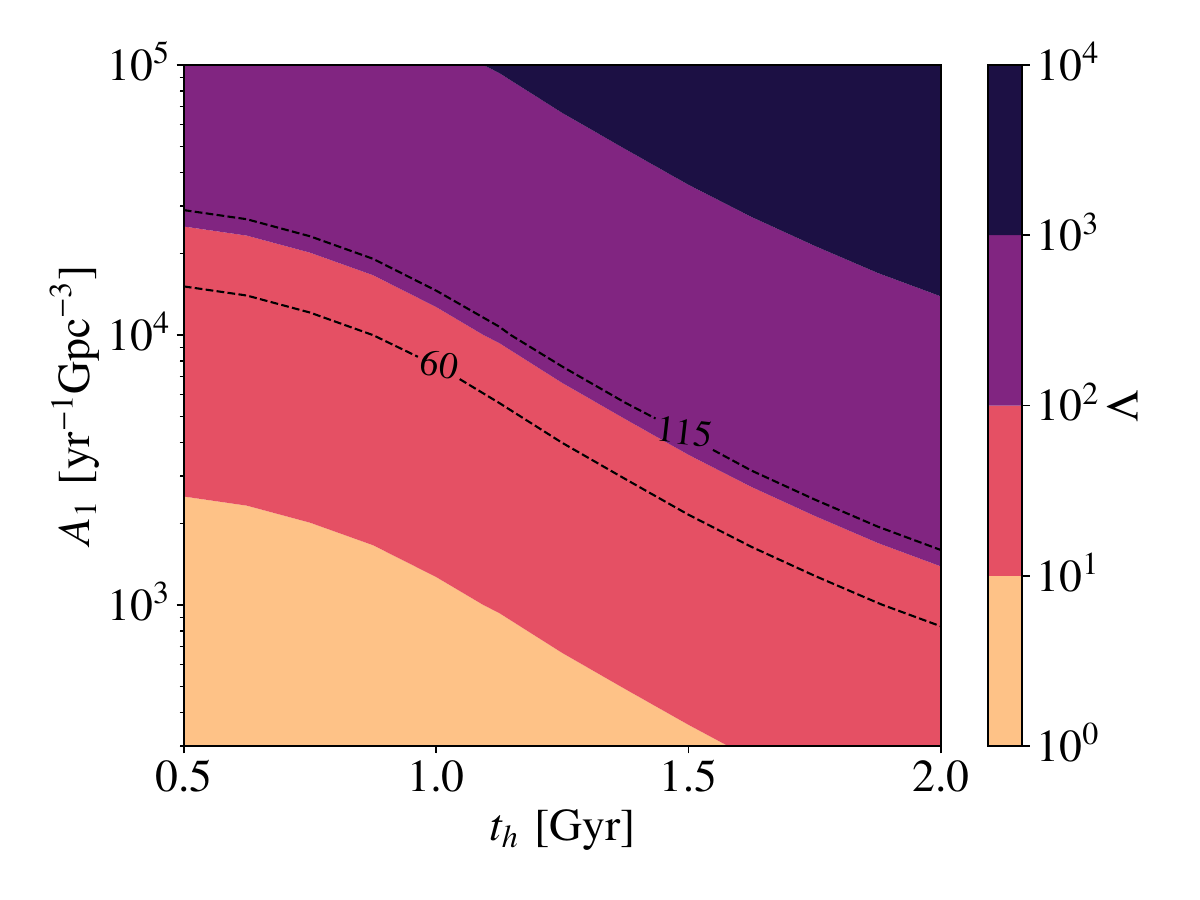}
    \caption{The expected number of detectable GW events $\Lambda$ as a function of BDS model parameters $A_1$ and $t_h$. We draw contours of $\Lambda = 60$ and 115 (number of GW detections in O1-O2-O3, including Poisson uncertainties; see text and Appendix~\ref{appendix:detectable_events_and _strongly_lensed_pairs_thresholds} for more details). The region outside the dashed contours is inconsistent with the observed events.}
    \label{fig:detectable_events}
\end{figure}
\begin{figure}
    \centering
    \includegraphics[width=\linewidth]{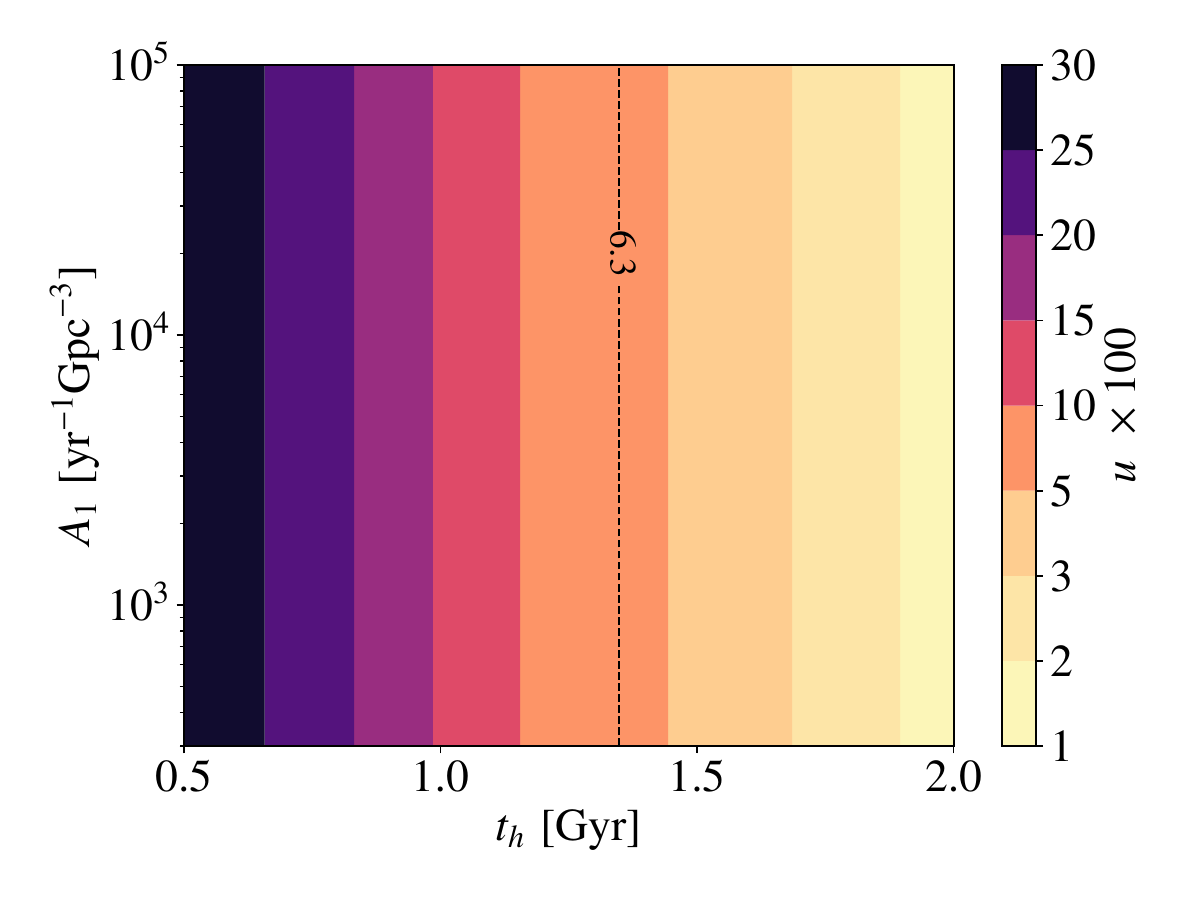}
    \caption{The expected percent of detectable strongly lensed pairs as a function of $A_1$ and $t_h$. The $3\sigma$ upper bound on the observed strong lensing fraction is $\sim 6.3\%$ (see Appendix~\ref{appendix:on_the_non_detection_of_strongly_lensed_pairs_until_o3b}). The region with a larger fraction than this bound is inconsistent with the data.}
    \label{fig:detectable_doubles_frac}
\end{figure}

\paragraph{Based on the Detectable Fraction of Strongly Lensed Pairs}~\label{subsec: result_discu_strongly_lensed_pairs}
We find the fraction of detectable strongly lensed pairs among the total detectable events as a function of $A_1$ and $t_h$, using the approach described in Sec.~\ref{subsec: strongly_lensed_pairs}. In Fig.~\ref{fig:detectable_doubles_frac}, the expected fraction predicted by the BDS model lies within the range $2- 30\%$. The $3\sigma$ upper bound on the observed strong lensing fraction estimated from O1-O2-O3 data is $\sim 6.3\%$ (see Appendix~\ref{appendix:on_the_non_detection_of_strongly_lensed_pairs_until_o3b}). The region of the parameter space where the predicted fraction exceeds the upper bound is inconsistent with the observations. This eliminates a large portion of the $A_1-t_h$ space, on the left of the bound.

\begin{figure}[t]
    \centering
    \includegraphics[width=\linewidth]{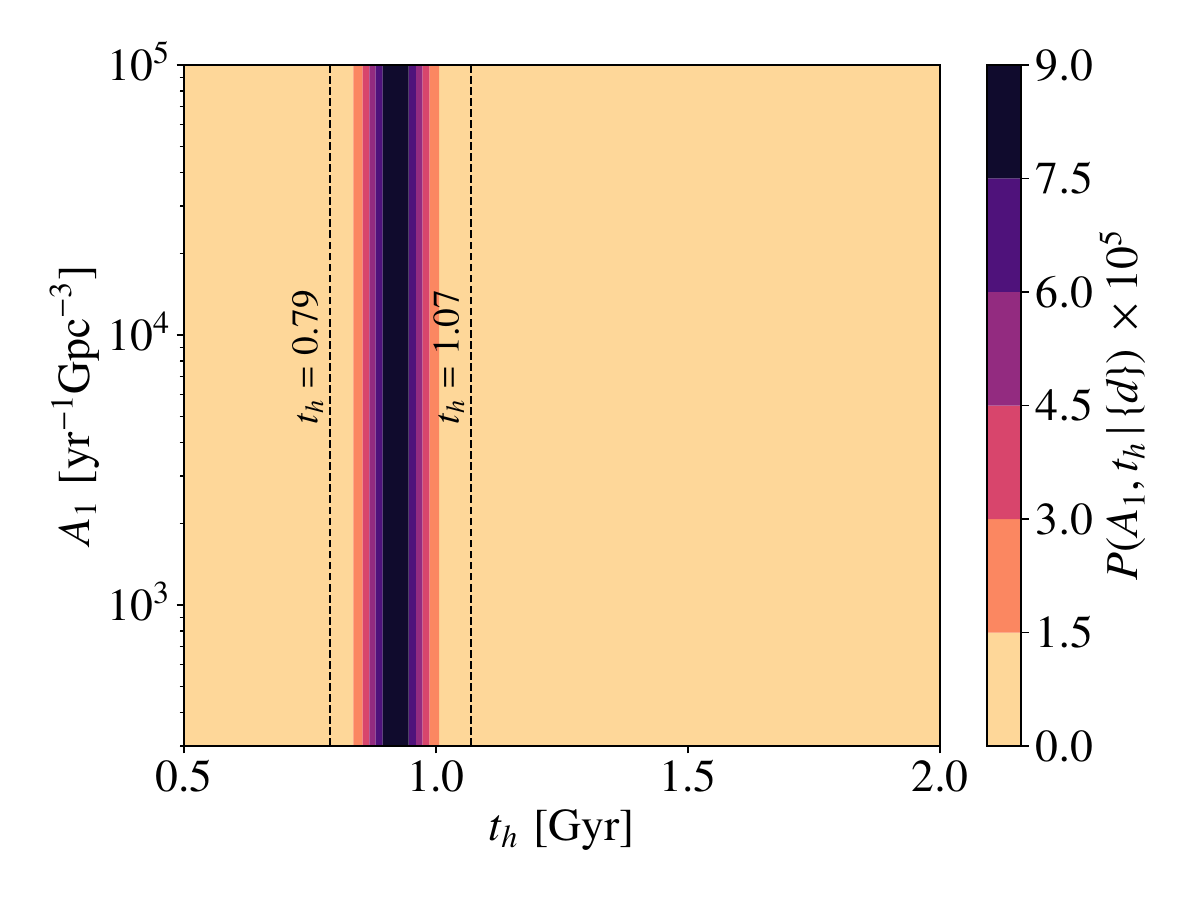}
    \caption{The posterior distribution of $A_1, t_h$ estimated from the consistency of the predicted distribution of the redshifted total mass and apparent luminosity distance with the corresponding posteriors of the observed events. The $3\sigma$ lower/upper bounds are marked by vertical dashed lines.}
    \label{fig:th_region_from_mass_dist}
\end{figure}
\begin{figure}[t]
    \centering
    \includegraphics[width=\linewidth]{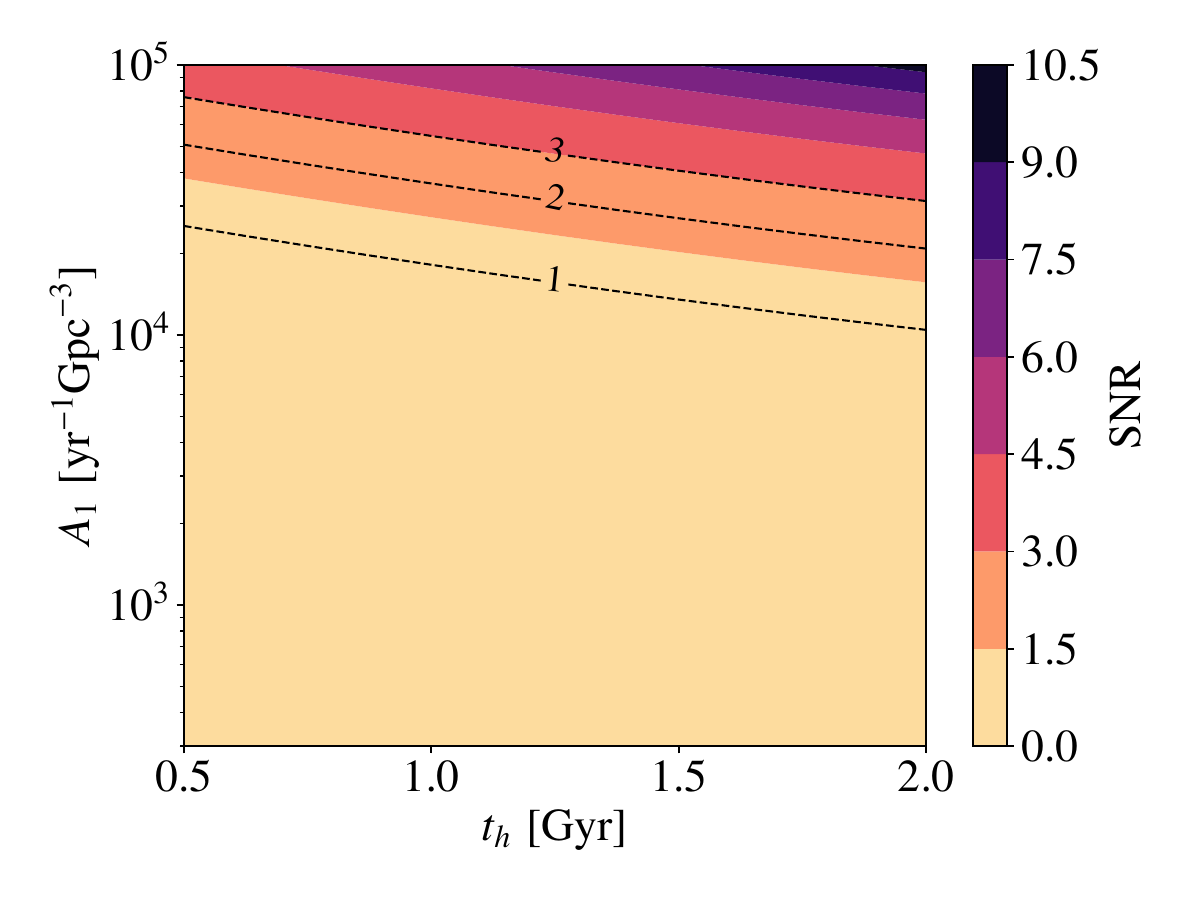}
    \caption{The SGWB SNR as a function of $A_1$ and $t_h$. We draw contours of SNR of 1, 2 and 3 (which correspond to a $1\sigma, \,2\sigma$ and $3\sigma$ detection of SGWB, respectively). The parameter space above an SNR of 3 can be rejected because of the non-detection of SGWB in current observations.}
    \label{fig:sgwb-snr}
\end{figure}

\paragraph{Based on the Distribution of Redshifted Masses and Apparent Luminosity Distance}~\label{subsec: result_discu_mass_distr}
We simulate the expected distribution $P_\mathrm{pop}(\theta|t_h)$ of the redshifted total mass and apparent luminosity distance corresponding to different values of $t_h$. We then compute the posterior of $t_h$ using Eqs.~\eqref{eq:th_likehood_from_mass_dist} and \eqref{eq:th_posteriors_bayes_theorem}, assuming a flat prior in $t_h$. The integrals in Eq.~\eqref{eq:th_likehood_from_mass_dist} are computed using Monte-Carlo integration as in \cite{Thrane:2018qnx, Mandel:2018mve}. We then compute the $3\sigma$ credible region of the posterior, which is shown in Fig.~\ref{fig:th_region_from_mass_dist}. The posterior peaks at $t_h \simeq 0.95 \, \mathrm{Gyr}$. This is the most favorable region in the parameter space for reproducing the observed distribution of $M^z$ and $\tilde{d_L}$. Note, however, that even this choice of parameters does not satisfactorily reproduce the observed distribution of redshifted total mass and apparent luminosity distance (see Appendix \ref{app:mass_dl_dist}).

\paragraph{Based on the Non-Detection of Stochastic GW Background}~\label{subsec: result_discu_sgwb}
Figure~\ref{fig:sgwb-snr} presents the SNR computed using the approach described in Sec.~\ref{subsec: sgwb_methods}. The SNR ranges from $\sim 0$ to $10.5$. We draw contours of SNR 1, 2 and 3. The increase of the SNR with $A_1$ can be attributed to the increase in the merger rate. Similarly, the SNR increases with $t_h$, although slowly, because of the increase in the number of mergers at low redshifts. In Fig.~\ref{fig:sgwb-snr}, the parameter space above the contour of SNR $3$, with $3\sigma$ confidence bound, can be rejected because we have not observed such a background.

We would like to particularly highlight the parameter pair $A_1 = 30000 \, \mathrm{yr}^{-1} \, \mathrm{Gpc}^{-3}$ and $t_h=1.25 \, \mathrm{Gyr}$, as reported in \cite{Diego:2021fyd}, which is said to offer a strong match between the GWTC-3 catalog~\citep{KAGRA:2021vkt, KAGRA:2021duu} and the BDS model. {In our analysis, this combination is inconsistent with key observables: an excess of detectable events, a surplus of strongly lensed pairs, and an amplified SGWB (see Appendix~\ref{appendix:stochastic_background}). In fact, we find no region in the $A_1 - t_h$ parameter space that is consistent with all the observational data -- each region is ruled out by one or more observations.

\section{Conclusion}\label{sec: conclusion}
\begin{figure}
    \centering
    \includegraphics[width=\linewidth]{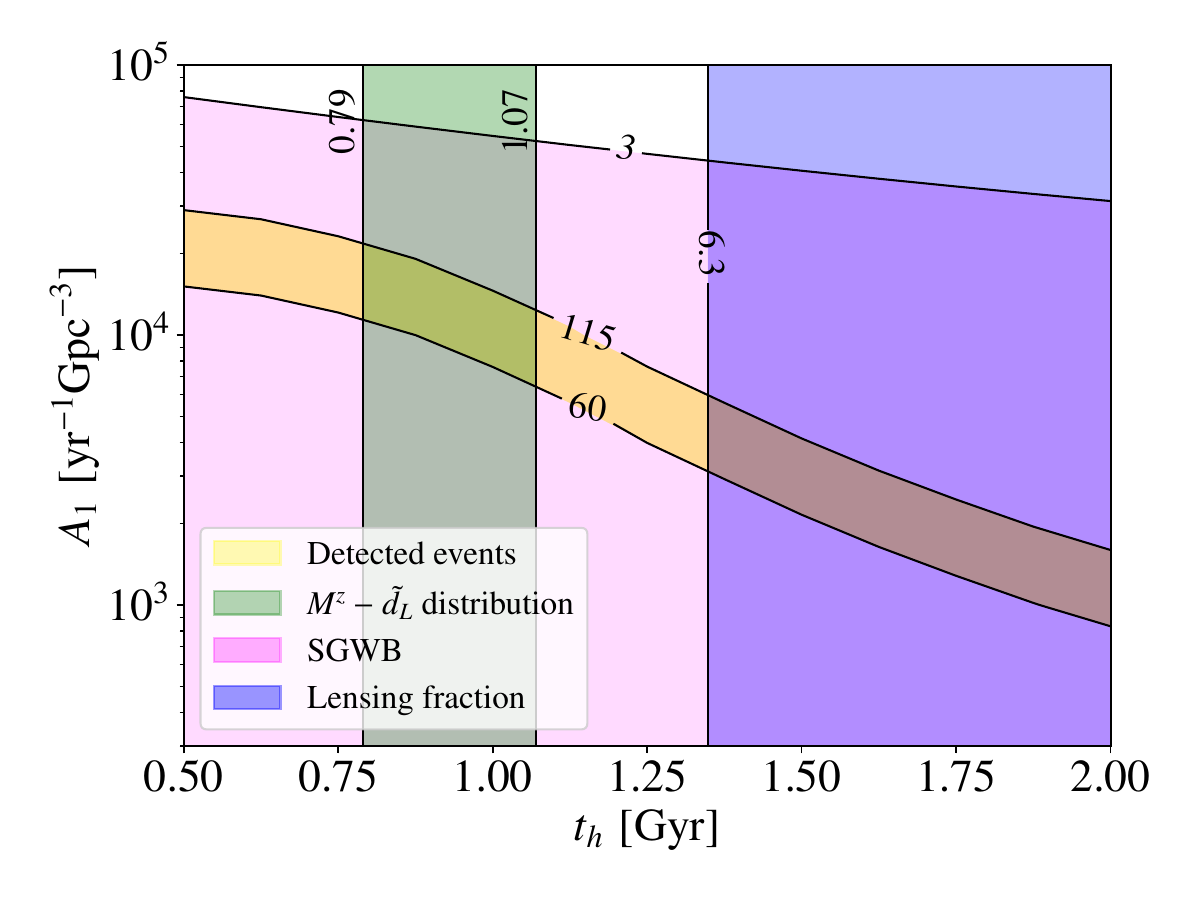}
    \caption{Viable regions of the BDS model parameter space under various criteria are colored: The yellow region is consistent with the total number of detected GW events (Fig.~\ref{fig:bds-merger-rate}), the green region is consistent with the observed redshifted total mass and apparent luminosity distance distribution (Fig.~\ref{fig:th_region_from_mass_dist}), the pink region is consistent with the non-detection of SGWB (Fig.~\ref{fig:sgwb-snr}). The blue region is consistent with the observed upper limit on the strong lensing fraction (Fig.~\ref{fig:detectable_doubles_frac}). No point in the parameter space is consistent with \emph{all} the four observables simultaneously.}
    \label{fig:combined_plot}
\end{figure}

We checked the consistency of the BDS model with different aspects of the observational data, namely, (1) the number of detectable GW events, (2) the non-detection of strongly lensed pairs, (3) the observed distribution of redshifted total masses and apparent luminosity distance, and (4) the SNR of the SGWB. Our findings indicate that different regions of the $A_1-t_h$ parameter space in the BDS model are inconsistent with different observations. When all these constraints are combined, no choice of parameters remains viable~(Fig.~\ref{fig:combined_plot}). 

In this study, we used the SIS lens model to compute the lensing optical depth and magnifications. The number of detectable GW events and the SGWB SNR are largely insensitive to the choice of lens model. However, the fraction of strongly lensed pairs as well as the distribution of the redshifted masses and apparent luminosity distance of detected events will have some dependence on the lens model that we assume. The modified elliptical Navarro-Frenk-White (NFW) profile that is considered by BDS~\citep{Diego:2019rzc} has optical depths comparable or lower than that of the SIS profile. This will not change our conclusions (see Appendix~\ref{appendix: choice_of_the_lens_model}).

The BDS model, with adjustable parameters $A_1$ and $t_h$, spans merger rate profiles predicted by a range of population synthesis studies (see the shaded region in figure \ref{fig:bds-merger-rate}). We believe that it would be difficult, if not impossible, to construct a BDS-like model that would successfully explain all the observations if the galactic BH mass distribution [Eq.~\eqref{eq:log-normal-mass-distr}] is assumed for the extragalactic binary BHs. 

Interestingly, supporting evidence for more massive BHs within our galaxy has emerged from recent Gaia observations~\citep{Gaia:2023fqm}. Lensing, while an intriguing phenomenon, has yet to be conclusively identified in GW observations~\citep{LIGOScientific:2021izm, LIGOScientific:2023bwz}. {Lensing is expected to feature prominently in future observing runs of LVK and with next-generation detectors} \citep{Li:2018prc, Barsode:2025agk, Maity:2025apt}, where lensing magnification could significantly affect the inference of the population properties of merging BH binaries. We are currently developing a formalism for BH population inference considering lensing effects, making use of the different aspects of the observational data that we consider in this \emph{Letter}.  

\section*{acknowledgments}
We thank Alorika Kar, Ankur Barsode, Koustav N. Maity, and other members of the ICTS Astrophysical Relativity Group for useful discussions. We acknowledge the support of the Department of Atomic Energy, Government of India, under project nos. RTI4019 and RTI4013. 
PA acknowledges support from the ICTP through the Associates Programme and from the Simons Foundation, whose grant (Record ID: SFI-MPS-T-Institutes-00012057, AD) also supported this work.
Numerical calculations reported in the paper were performed on the Alice computing cluster at ICTS-TIFR. {This material is based upon work supported by NSF's LIGO Laboratory, which is a major facility that is fully funded by the National Science Foundation.}

\software{Analyses in this \emph{Letter} made use of Numpy~\citep{Harris:2020xlr}, Scipy~\citep{Virtanen:2019joe}, Astropy~\citep{Astropy:2013muo}, Ipython~\citep{Perez:2007emg}, PyCBC~\citep{Usman:2015kfa, Nitz:2018rgo, 2020zndo...3596447N, Davies:2020tsx}, GWpy~\citep{Macleod:2021goi}, HMF~\citep{Murray:2013qza}, cogwheel~\citep{Roulet:2022kot}. For plotting, we have used Matplotlib~\citep{Hunter:2007ouj}.}

\appendix
\setcounter{section}{0}
\gdef\thesection{\Alph{section}}

\section{Constraints on detectable events}\label{appendix:detectable_events_and _strongly_lensed_pairs_thresholds}
\begin{figure}
    \centering
    \includegraphics[width=\linewidth]{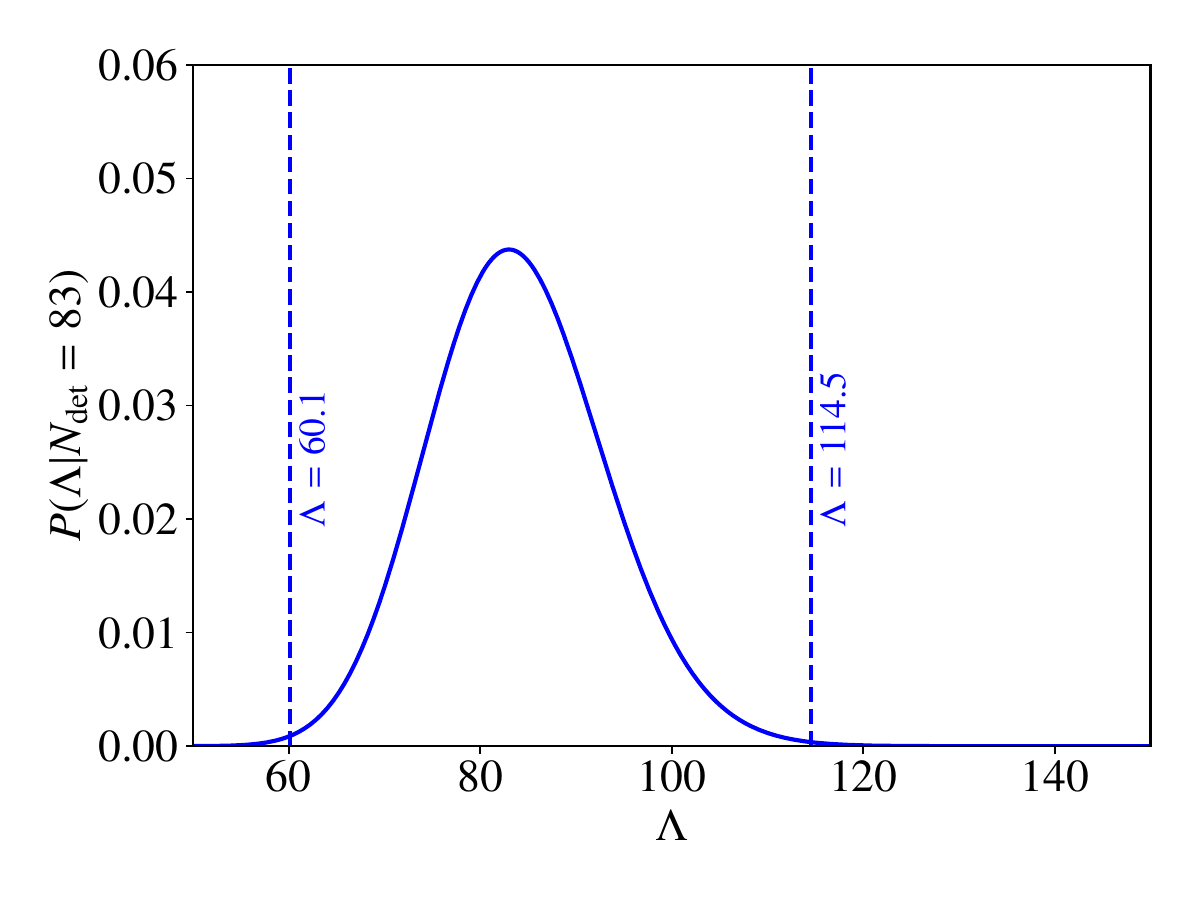}
    \caption{The posterior on the Poisson mean of the distribution of the number of detectable events. The vertical lines at $\Lambda \sim 60$ and $\Lambda \sim 115$ enclose the $3\sigma$ bounds on the Poisson mean $\Lambda$.}
    \label{fig:posterior_on_lambda}
\end{figure}

We assume that the number $N_\mathrm{det}$ of detected GW events follows a Poisson distribution with mean $\Lambda$. Given $\Lambda$, the likelihood of detecting $N_\mathrm{det}$ events is $P(N_\mathrm{det}|\Lambda)=\Lambda^{N_\mathrm{det}}\exp(-\Lambda)/N_\mathrm{det}!$. From O1-O2-O3 data, $N_\mathrm{det} = 83$ (considering events with median primary and secondary masses $\geq 5 M_\odot$)}. Assuming a prior $P(\Lambda)$, we can estimate the posterior distribution of $\Lambda$ using Bayes theorem: 
\begin{equation}
    \label{eq:posterior_on_lambda}
    P(\Lambda | N_\mathrm{det}) = \frac{P(N_\mathrm{det}|\Lambda)P(\Lambda)}{\int d\Lambda\,P(N_\mathrm{det}|\Lambda)P(\Lambda)}.
\end{equation}
We show the posterior in figure~\ref{fig:posterior_on_lambda}, assuming a flat prior on $\Lambda$. The vertical lines at $\sim 60$ and $\sim 115$ enclose a $3\sigma$ credible region of the posterior.

\section{Stochastic GW background}\label{appendix:stochastic_background}
\begin{figure}
    \centering
    \includegraphics[width=0.5\textwidth]{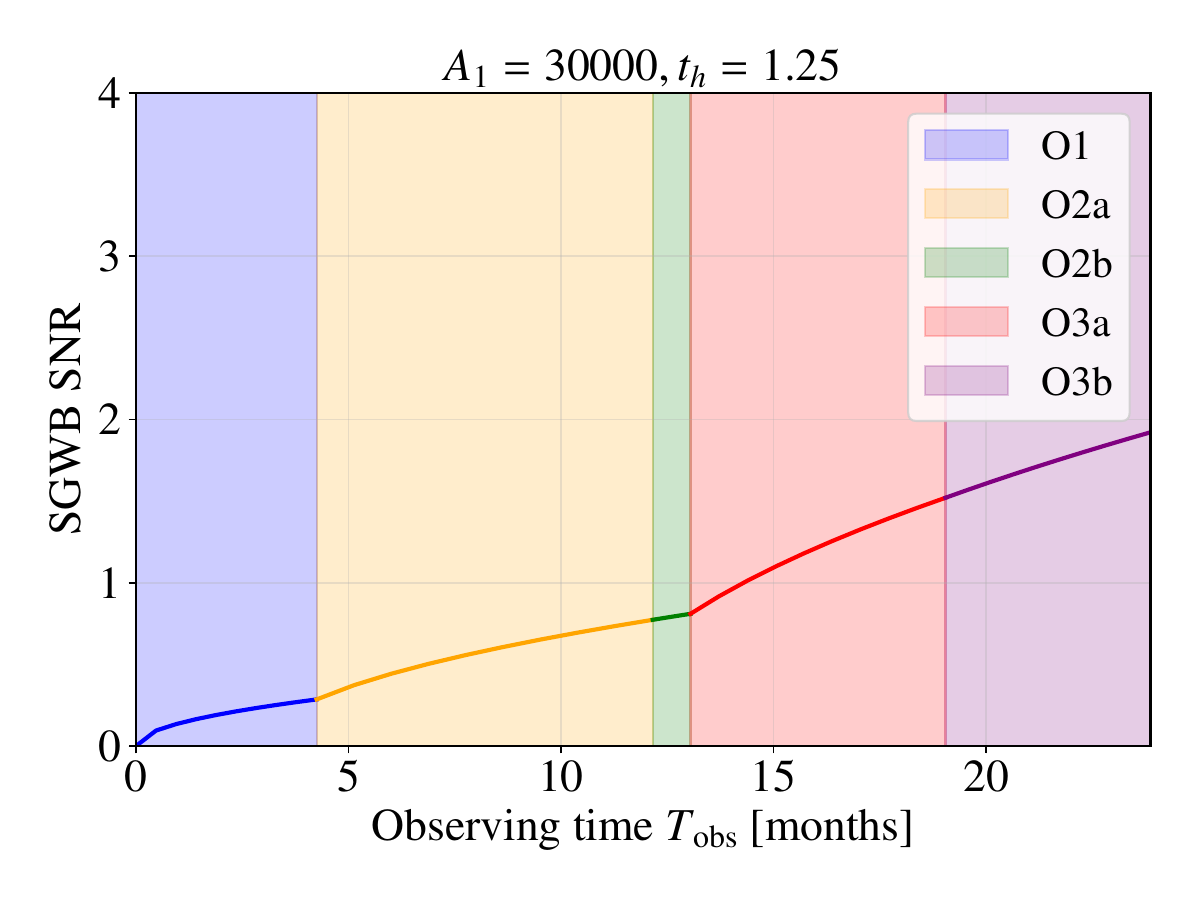}
    \caption{The accumulated SNR of the SGWB  through O1, O2 and O3 observing runs for $A_1 = 30000 \, \mathrm{yr}^{-1} \, \mathrm{Gpc}^{-3}$ and $t_h = 1.25 \, \mathrm{Gyr}$. The SNR exceeds the threshold of 1 in the O3a run, which, when combined with the non-detection of SGWB, rules out this choice of BDS model parameters.}
    \label{fig:SGWB_SNR_30000_1.25}
\end{figure}

\cite{Mukherjee:2020tvr} also investigated the detectability of SGWB using the BDS model, assuming $A_r = 2400 \, \mathrm{yr}^{-1} \, \mathrm{Gpc}^{-3}$ and $t_h = 1.5 \, \mathrm{Gyr}$ and concluded that the non-detection in O1-O2 data is consistent with the BDS model. In Fig.~\ref{fig:SGWB_SNR_30000_1.25}, we show the evolution of the SGWB SNR during O1, O2, and O3 observing runs for $A_1 = 30000 \, \mathrm{yr}^{-1} \, \mathrm{Gpc}^{-3}$ and $t_h = 1.25 \, \mathrm{Gyr}$ that was found optimal by~\cite{Diego:2021fyd}. We observe that the SNR crosses 1 in O3a, and comes close to 2 in O3b. The fact that we have not detected the SGWB signal in the LVK observations disfavors this choice of parameters. 

\section{Choice of the lens model}\label{appendix: choice_of_the_lens_model}
\begin{figure}
    \centering
    \includegraphics[width=\linewidth]{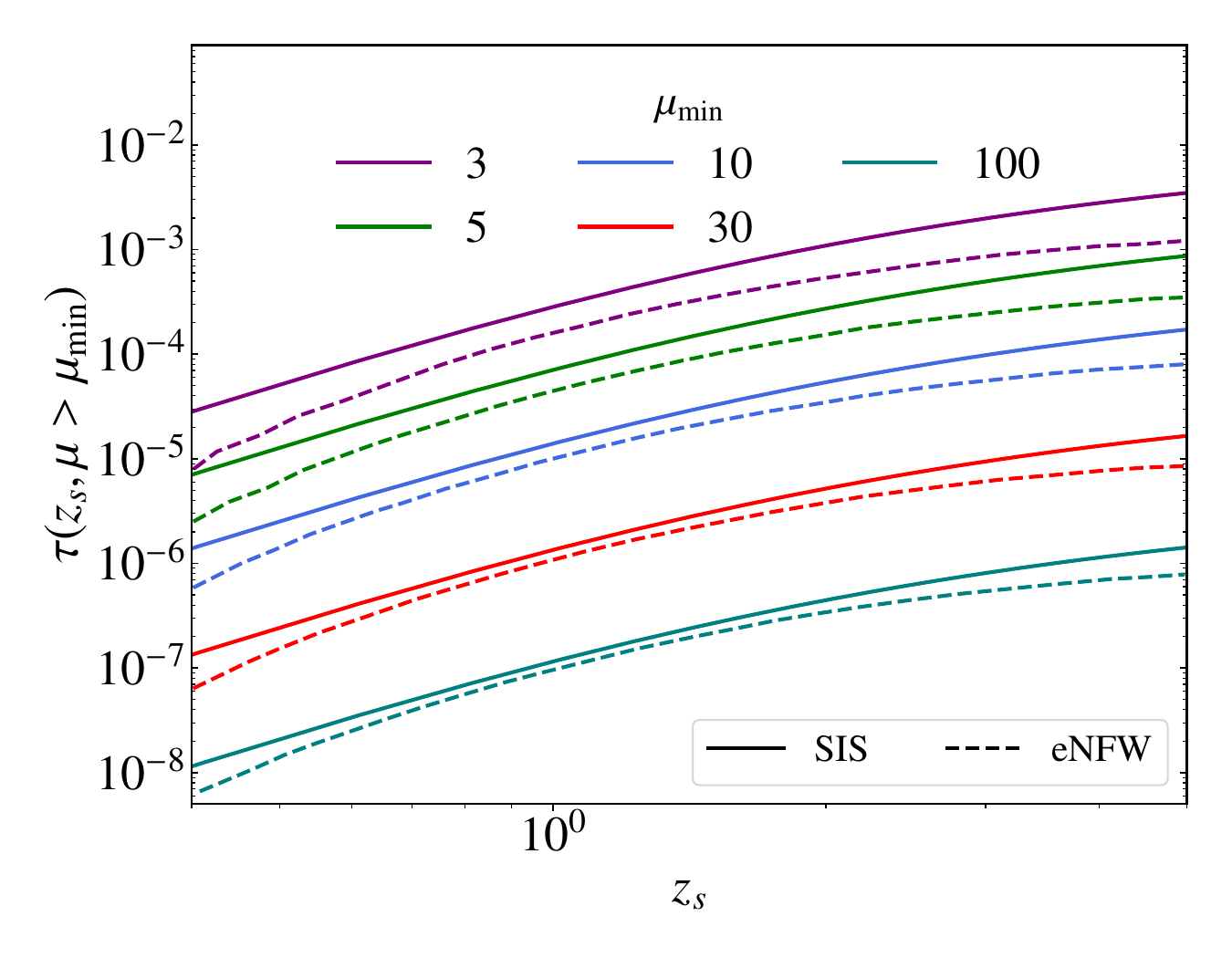}
    \caption{Comparison of the lensing optical depth of SIS lenses that we consider (solid lines) with that of the modified elliptical NFW lenses considered by BDS (dashed lines). The horizontal axis shows the source redshift $z_s$, while each line corresponds to the fraction of sources magnified with magnification greater than $\mu_\mathrm{min}$. The optical depth of the BDS lens model is comparable to or slightly lower than that of the SIS model that we use.} 
    \label{fig:opt_depth_SIS_eNFW}
\end{figure}

Following~\cite{Jana:2022shb, Jana:2024uta}, we have used the SIS lens model to compute the optical depth in Eq.~\eqref{eq:dPdy}
\begin{equation}
\tau(z_s) = \int_0^{z_s} d z_\ell \int_{\sigma_\mathrm{min}}^{\sigma_\mathrm{max}} d\sigma \frac{dV_c}{dz_\ell} ~ \frac{dN_\ell}{dV_c d\sigma} ~ \frac{\pi r_E^2}{4 \pi D_\ell^2}, 
\label{eq:opt_depth}
\end{equation}
where $z_\ell$ is the lens redshift and $\sigma$ is the velocity dispersion of the lens, ${dV_c}/{dz_\ell}$ is the differential comoving volume, ${dN_\ell}/{dV_c d\sigma}$ is the comoving number density of lenses with velocity dispersion $\sigma$ at redshift $z_\ell$, $r_E$ is the Einstein radius of a lens at $z_\ell$ with velocity dispersion $\sigma$, and $D_l$ is the angular diameter distance to the lens. The number density of lenses is computed as 
\begin{equation}
\frac{dN_\ell}{dV_c d\sigma}  = \frac{dN_\ell}{dV_c dM} \frac{dM}{d\sigma}, 
\end{equation}
where the first term denotes the comoving number density of dark matter halos with mass $M$ at redshift $z_\ell$ given by the halo mass function model of \cite{Behroozi:2012iw}. We use the Jacobian ${dM}/{d\sigma} = 3M/\sigma$ to convert a dark matter halo of mass $M$ into an SIS lens of velocity dispersion $\sigma$. The limits $\sigma_\mathrm{min}$ and $\sigma_\mathrm{max}$ in Eq.~\eqref{eq:opt_depth} corresponds to the lightest and heaviest dark matter halos that we consider, with masses $10^8 M_\odot$ and $10^{15} M_\odot$. Thus, we consider both galaxy and cluster scale lenses. See Sec.~2.2.1~\cite{Jana:2024uta} or the Supplementary Material of~\cite{Jana:2022shb} for more details. 

The lensing optical depth $\tau(z_s)$ and magnification distributions depend on the lens model assumptions. Singular isothermal ellipse lenses have an optical depth comparable to or lower than that of the SIS model~\citep{Huterer:2004jh}, resulting in a similar or reduced number of strongly lensed events. The NFW model will have much smaller optical depths~\citep{2012arXiv1206.4919S,Brando:2024inp,Vujeva:2025kko}. 

BDS uses a modified elliptical NFW model with a steeper central region that aims to model the baryonic contribution~\citep{Diego:2019rzc}. The optical depth of this model is comparable to or slightly lower than the SIS model that we use (see Fig.~\ref{fig:opt_depth_SIS_eNFW}). It will be even harder for the BDS lens model to produce the required lensing magnification to explain the observed high-mass BHs. Hence, we do not expect this to alter our conclusions.

\section{Constraints on the lensing fraction}\label{appendix:on_the_non_detection_of_strongly_lensed_pairs_until_o3b}

\begin{figure}
    \centering
    \includegraphics[width=\linewidth]{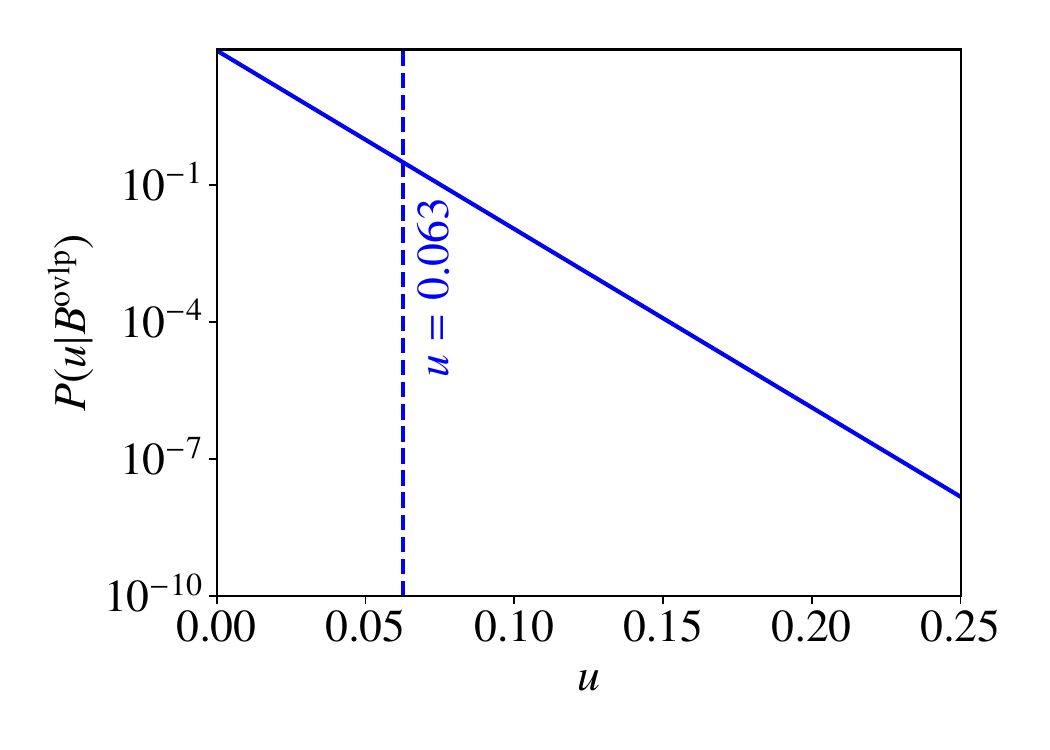}
    \caption{The posterior and $3\sigma$ upper bound on the strong lensing fraction given the $B^\mathrm{ovlp}$ values of LVK event pairs. % for $t_h = 1.25$ Gyr. 
    We get a $3\sigma$ upper bound of $6.3\%$.}
    \label{fig:strong_lensing_fraction_bound_80000_1p25}
\end{figure}
\begin{figure}
    \centering
    \includegraphics[width=\linewidth]{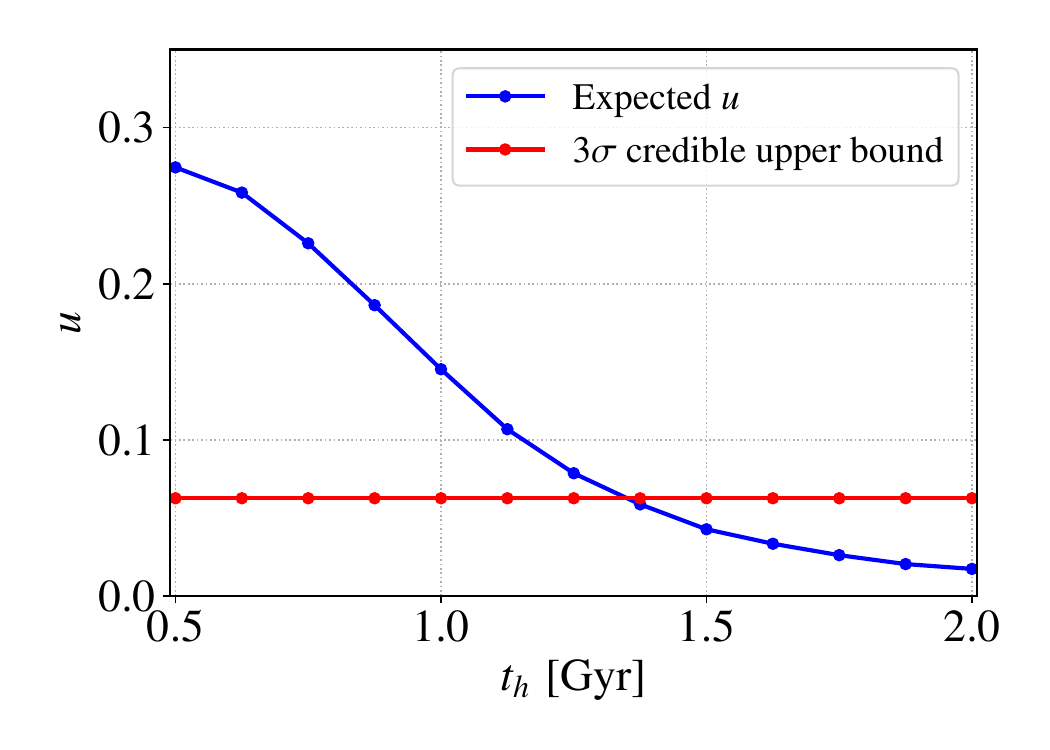}
    \caption{The fraction $u$ of detectable strong lensing events predicted by the BDS model as a function of $t_h$, and the observed $3\sigma$ upper limit. For $t_h < 1.4$ Gyr, the predicted strong lensing fraction is greater than the observed upper bound.}
    \label{fig:expected_and_upper_bound_on_strong_lensing_fraction}
\end{figure}

Figure \ref{fig:strong_lensing_fraction_bound_80000_1p25} shows the posterior and $3\sigma$ upper bound on the observed strong lensing fraction $u$ given the set of $B^\mathrm{ovlp}$ values from O1-O1, O2-O2, O3a-O3a, and O3b-O3b pairs. 
The lensing fraction is consistent with being zero, with a $3\sigma$ upper limit of 6.3\%. 

Figure~\ref{fig:expected_and_upper_bound_on_strong_lensing_fraction} shows the predicted lensing fraction by the BDS model as a function of the parameter $t_h$, as well as its $3\sigma$ upper limits from the data. Values of $t_h$ predicting a lensing fraction that is larger than the observed upper bound can be ruled out. 

\section{Distribution of redshifted total mass and apparent luminosity distance} \label{app:mass_dl_dist}
\begin{figure}[h]
    \centering
    \includegraphics[width=\linewidth]{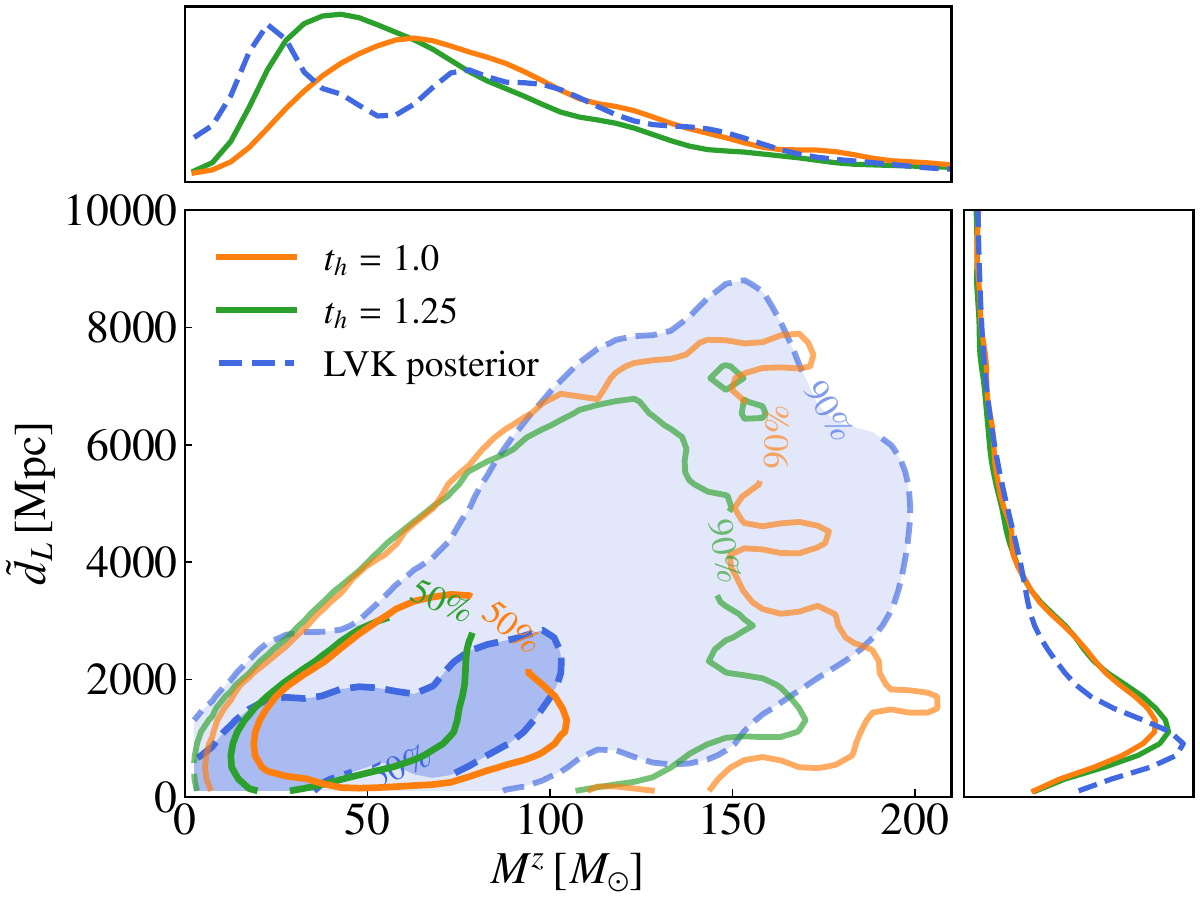}
    \caption{
    The blue contours show the distribution of the redshifted total mass and apparent luminosity distance of all the detected events inferred from the data. The orange and green contours correspond to the prediction of the BDS model with parameters $t_h = 1.0 \, \mathrm{Gyr}$ (close to the best fit value in our analysis) and $t_h = 1.25 \, \mathrm{Gyr}$ (the optimal choice as claimed by BDS). None of the model choices are particularly good in reproducing the observed distribution.}
    \label{fig:Mz_dLapp_contour_plot}
\end{figure}
Here we compare distribution of the redshifted total mass and apparent luminosity distance of detected events that we infer from the data (all events combined), with the same predicted by BDS model. Figure~\ref{fig:Mz_dLapp_contour_plot} shows the inferred distribution in the redshifted total mass $M^z$ and the apparent luminosity distance $\tilde{d}_L$, along with the model prediction corresponding to $t_h = 1.25 \, \mathrm{Gyr}$ (the optimal choice as claimed by BDS) as well as $t_h = 1.0 \, \mathrm{Gyr}$ (close to the best fit value in our analysis). It can be seen that none of the model choices are particularly good in reproducing the observed distribution. In particular, the BDS model is unable to reproduce the bimodal nature of the mass distribution, unlike what is claimed in~\cite{Diego:2021fyd}. The Jensen-Shannon divergence $D_\mathrm{JS}$~\citep{Lin1991DivergenceMB} between the LVK posterior and BDS model is $0.32 ~ (0.27)$ for $t_h = 1 ~ (1.25)$. For reference, two identical distributions have $D_\mathrm{JS}=0$, while two completely unrelated distributions have $D_\mathrm{JS}=1$.
\bibliography{references}{}
\bibliographystyle{aasjournal}

\end{document}